\UseRawInputEncoding
\documentclass[12pt]{amsart}
\usepackage{amsmath}
\usepackage{amsthm}
\usepackage{amsfonts}
\usepackage{graphicx}
\usepackage{amssymb} 

\newenvironment{nouppercase}{
  
  \renewcommand{\uppercasenonmath}[1]{}}{}

\begin{document}

\title[The ground state of reduced Yang-Mills theory]
{The ground state of reduced Yang-Mills theory}
\author[Jens Hoppe]{Jens Hoppe}
\address{Braunschweig University, Germany \& IHES, France}
\email{jens.r.hoppe@gmail.com}

\begin{abstract}
For the simplest membrane matrix model 
(corresponding to reduced 3 dimensional $SU(2) $ Yang Mills theory) the form of the ground state wave function is given.
\end{abstract}

\begin{nouppercase}
\maketitle
\end{nouppercase}
\thispagestyle{empty}
\noindent
Relativistic membranes in $D$-dimensional space-time can be described by a $ SU(N \rightarrow \infty)$  Matrix Model \cite{1} involving $d=D-2$ hermitean $N \times N$ matrices $ X_i $ that interact via a potential given by (minus) the trace of the sum of squares of commutators $\left[ X_i,X_j\right]$.
In the simplest case, $d=2=N$, the potential becomes the square of the area spanned by (the origin and) two particles in $\mathbb{R}^3$ \cite{2}. At the same time (\cite{3}; as far as I remember, also Feynman investigated in some detail the 2+1 dimensional case \cite{4}) the models were considered as reductions of d+1 dimensional Yang-Mills theory. 

Simon \cite{5} and L\"uscher \cite{6} proved that the spectrum of the quantum mechanical models is purely discrete (related results, communicated by H.Kalf in 1982, had been proven by Rellich \cite{7}).

In order to find the structure of the ground state $\psi$ of \cite{2}
\begin{equation}\label{eq1} 
H = \frac{1}{2}(-\vec{\nabla}_1^2-\vec{\nabla}_2^2)+(\vec{q_1} \times \vec{q_2})^2
\end{equation}
take $\psi$ to be a function of the two $SO(3)\times SO(2)$-invariant variables 
\begin{equation}\label{eq2} 
U:=\frac{1}{2}(\vec{q_1}^2 + \vec{q_2}^2) \quad V:=\frac{1}{2}(\vec{q_1} \times \vec{q_2})^2 .
\end{equation}
Up to some numerical constants one has (for invariant integrands)
\begin{equation}\label{eq3} 
\int d^3q_1 d^3 q_2  \sim \int dU dV 
\sim \int dx_1 dx_2 \overbrace{(x_1 x_2 (x_2^2-x_1^2))}^{\rho(x_1,x_2)}
\end{equation}
where $x_2^2 \geq x_1^2 \geq 0$ are the eigenvalues of the positive semi definite matrix

\begin{equation}\label{eq4} 
B:= \begin{pmatrix}
    \vec{q}_1^2 & \vec{q}_1 \cdot \vec{q}_2\\
   \vec{q}_1 \cdot \vec{q}_2 & \vec{q}_2^2
   
\end{pmatrix}
\end{equation}
whose determinant is $x_1^2 \cdot x_2^2 = (\vec{q_1} \times \vec{q_2})^2$ .

While the resulting
\begin{align}\label{eq5}
\tilde{H}  & = \frac{1}{2}(-\partial_{x_1}^2-\partial_{x_1}^2 + x_1^2 x_2^2 + 2\tilde{V}) \nonumber \\
2\tilde{V} & =  \frac{1}{2}(\partial_{11}^2+\partial_{22}^2)(ln \rho) +\frac{1}{4} \frac{\rho_1^2 + \rho_2^2}{\rho^2}  \\
& =-(\frac{1}{4}(\frac{1}{x_1^2}+\frac{1}{x_2^2})+\frac{x_1^2+x_2^2}{(x_2^2-x_1^2)^2})=\frac{-4}{r^2\sin^2 \theta} \nonumber,
\end{align}
acting on $\tilde{\psi}:= \sqrt{\rho}\psi$ (i.e. vanishing at $x_1=0$ and for $x_1=x_2$), has a nice interpretation as a generalized Calogero Moser system, the variables in (2), $C^{\infty}$-functions (resp. real-analytic) on $\mathbb{R}^6$, allow a power series expansion for $\psi$.
Using
\begin{equation}\label{eq6} 
(\nabla U)^2=2U, \quad (\nabla V)^2=4 UV, \quad \nabla U \cdot \nabla V=4V
\end{equation}
one gets for the kinetic energy (starting with $\frac{1}{2}(\nabla \psi)^2)$ and from now on writing u,v 
for U,V)
\begin{equation}\label{eq7}
T=\int_0^{\infty}du \int_0^{\frac{u^2}{2}}dv (u \psi_u^2+2uv\psi_v^2+4v\psi_u\psi_v) .
\end{equation}
Defining 
\begin{equation}\label{eq8}
w:=\frac{2v}{u^2}=\frac{4 x_1^2 x_2^2}{(x_1^2+x_2^2)^2}=\sin^2\frac{\theta}{2} \in \left[0,1\right]
\end{equation}
removes the curvilinear boundary in (7), and easily leads to
\begin{equation}\label{eq9}
\begin{split}
3u\phi_u+u^2(\phi_{uu}-\phi_u^2)+4(1-2w)\phi_w+4w(1-w)(\phi_{ww}-\phi_w^2)\\
+\frac{1}{2}wu^3=Eu
\end{split}
\end{equation}
when writing $\psi$ as minus the exponential of

\begin{equation}\label{eq10}
\begin{split}
\phi= Au+Bv+C\frac{u^2}{2}+Duv+F\frac{v^2}{2}+\frac{\mu u^3}{3} + ...
=\\
Au+ B \frac{w u^2}{2}+C \frac{u^2}{2}+D \frac{u^3 w}{2}+F \frac{w^2 u^4}{8}+ \mu \frac{u^3}{3}+...
\end{split} .
\end{equation}
The first few resulting relations are
\begin{equation}\label{eq11}
\begin{array}{ll}
3A = E, & 4C+2B-A^2 = 0\\[0.25cm] 7D+1 = 4AB, & 5\mu = 2AC-2D; 
\end{array} 
\end{equation}
hence
\begin{equation}\label{eq12}
\psi=e^{-\frac{Eu}{3} -B\frac{wu^2}{2} -\frac{(uE)^2}{2}\big(\frac{1}{36} - \frac{B}{2E^2}\big) + \frac{1}{14}  w (uE)^3 \big ( \frac{1}{E^3} - \frac{4B}{3E^2} \big) - \frac{(Eu)^3}{3}\big[ \big(\frac{1}{270}- \frac{B}{7E^2} \big) + \frac{2}{35E^3} \big]+ \ldots} .
\end{equation}
While the pure $(E u)^n$ terms (i.e. those with B=0, and without $1/E^3$) correspond to the expansion of $\ln(\frac{2}{Eu}J_2(2 \sqrt{E u}))$, where
\begin{equation}\label{eq13} 
\frac{1}{Eu}J_2(2 \sqrt{E u})=\sum_{k=0}^{\infty} (Eu)^k \frac{(-1)^k}{k! (k+2)!}
\end{equation}
is ($\frac{1}{Eu}$ times) a Bessel function corresponding to the regular SO(6) invariant solution of the free Schr\"odinger equation $-\frac{1}{2}\Delta^2 \psi = E \psi$ 

(note that, therefore, without the terms originating from the term $\frac{1}{2} wu^3$ in (9) the power-series would definitely have only a finite radius of convergence, equal to the first zero of (13)),
 there will be precisely one value of E (and one for B) for which (12) (when including the higher order terms) will be square integrable. A good approximate value of E (independent of all other considerations) can be obtained from
\begin{equation}\label{eq14} 
\langle H \rangle = 
\frac{1}{2}\int_0^{\infty}du \int_0^1 dw ( u^3 \psi_u^2 
+4uw(1-w)\psi_w^2+\frac{u^4 w}{2}\psi^2 )
\end{equation}
( where here $\psi$ is assumed to be normalized such that
$\frac{1}{2}\int du \int dw u^2 \psi^2$ equals 1 ) by choosing suitable trial-functions. 

The non-trivial observation is the form of the ground state, (12).

\medskip

\noindent 
Concluding remarks:
\medskip

1) (12), and the argument leading to it (in the case of reduced 3+1 dimensional Yang-Mills theory, the natural third variable, together with U and V, is the determinant of the 3 position vectors, which has scaling dimension 3 - so together with 2U/=sum of squares:quadratic/ and 2V/sum of squares of cross-products:quartic/ has the same dimension, 9, than the original position-space), should also be useful in the context of the important question of zero-energy states in supersymmetric matrix models (cf. e.g. \cite{8} \cite{9} \cite{10} \cite{11}, and references therein)
\medskip


\noindent 2) (10)
also arises if expanding
\begin{equation}\label{eq15} 
\psi = \sum_{\ell , n = 0}^{\infty} \psi_{\ell n} (r) \ P_{\ell} (\cos \theta = 1-2 w) \, c_{\ell n} \, ,
\end{equation}
where
\begin{equation}\label{eq16} 
\psi_{\ell n} (r) = 4 \cdot 
\sqrt{\frac{(2\ell +1) n!}{(n + 4\ell +2)!}}\ \exp({-r^2/2}) \, r^{4\ell} \, L_n^{4\ell +2} (r^2) \, ,
\end{equation}
the $L_n^{4\ell +2}$ being generalized Laguerre-polynomials, are suitable eigenfunctions of a 6-dimensional harmonic oscillator,
\begin{equation}\label{eq17}
\frac12 \left\{ - \psi''_{\ell n}  - \frac5r \psi'_{\ell n} + \frac{16 \ell (\ell + 1)}{r^2} \psi_{\ell n} + r^2 \psi_{\ell n} \right\} = E_{\ell n} \, \psi_{\ell n} ,
\end{equation}
and the $P_{\ell} (\cos \theta)$ are the standard Legendre polynomials, satisfying
\begin{equation}\label{eq18}
\frac1{\sin \theta} \partial_{\theta} \, \sin \theta \, \partial_{\theta} \, P_{\ell} (\cos \theta) = - \ell (\ell +1) P_{\ell}
\end{equation}
(note that due to $r^{4\ell}$ each power of $w$ necessarily comes with a factor of $u^2$).
While the matrix-elements
\begin{equation}\label{eq19}
\widetilde V_{n \ell , n' \ell'} = \frac1{16} \int_0^{\infty} r^5 dr \int_0^{\pi} \sin\theta \, d \theta \, P_{\ell} \, \psi_{n \ell} \left( \frac{r^4}{16} (1- \cos \theta) - r^2/2 \right) P_{\ell} \, \psi_{\ell' n'}
\end{equation}
(zero unless $(n-n')$ and $(\ell - \ell')$ are small) can be calculated in closed form, the question of lowest eigenvalue of
\begin{equation}\label{eq20}
\delta_{nn'} \, \delta_{\ell \ell'} (4\ell + 2n + 3) + \widetilde V_{n\ell , n'\ell'}
\end{equation}
is of course non-trivial.

\medskip
\noindent 3) (3) easily follows on dimensional grounds, when combined
with a few simple observations:
the Jacobian determinant that one gets from the singular value decomposition
\begin{equation}\label{eq21}
Q = (\vec q_1 \, \vec q_2) = S \begin{pmatrix} x_1 &0 \\ 0 &x_2 \\ 0 &0 \end{pmatrix} R^T
\end{equation}
is a quartic polynomial in $x_1$ and $x_2$ that vanishes when either of the two is zero (the 6 vectors in the corresponding matrix in that case become linearly dependent as the first, or second, column vector of S, having unit length, when differentiated wrt the 3 different angle-variables S depends on can not produce 3 linearly independent three dimensional vectors); as the quartic polynomial on the rhs of (3) 
is the Jacobian determinant of changing variables to U and V, the middle part of (3) can not have an extra w-dependent factor - the only one that would still give a quartic polynomial, the inverse square-root of w, would 
result in the difference of 4-th powers, which would not vanish at $x_1=0$.


\medskip

\textbf{Acknowledgement}:

I would like to thank T.Anous, V.Bach, T.Damour, T.Fischbacher, J.Fr\"ohlich, M.Hanada, M.Hynek, M.Kontsevich, I.Kostov, K.Kozhasov, K.Kim, I.Nafiz, D.Serban and T.Turgut for discussions

\end{document}